\documentclass[fleqn,11pt,twoside]{article}

\usepackage{amsmath,amssymb,amsthm}
\usepackage{accents}
\usepackage{tikz}
\usepackage{tikz-3dplot}
\usetikzlibrary{positioning,calc}
\def \h#1{\widehat{#1}}
\def \t#1{\widetilde{#1}}
\def \xht#1{\widehat{\widetilde{#1}}}
\def \xb#1{\overline{#1}}
\def \xbt#1{\overline{\widetilde{#1}}}
\def \xbh#1{\overline{\widehat{#1}}}
\def \xbht#1{\overline{\widehat{\widetilde{#1}}}}
\def \dd{\delta}
\def \sn {sn}
\def \ar{\alpha}
\def \bb{\beta}
\newcommand{\wh}[1]{\widehat{#1}}
\newcommand{\wt}[1]{\widetilde{#1}}
\newcommand{\wb}[1]{\overline{#1}}
\newcommand{\bse}{\begin{subequations}}
\newcommand{\ese}{\end{subequations}}
\newcommand{\nn}{\nonumber}
\newtheorem{prop}{Proposition}[section]

\makeatletter
\newcommand{\copyrightnote}[2]{{\renewcommand{\thefootnote}{}
 \footnotetext{\small\it
\begin{flushleft}
 \copyright \ #1   #2  
\end{flushleft}}}}

\newcommand{\Name}[1]{\begin{flushleft}
                       \LARGE \bf #1
                       \end{flushleft}\vspace{-3mm}}

\newcommand{\Author}[1]{\begin{flushleft}
                       \it #1 \end{flushleft}}

\newcommand{\Address}[1]{\begin{flushleft}
                       \it #1 \end{flushleft}}

\newcommand{\Date}[1]{\begin{flushleft}
                      \small  \it #1 \end{flushleft}}

\newcommand{\evenhead}{Author \ name}
\newcommand{\oddhead}{Article \ name}

\renewcommand{\@evenhead}{
\hspace*{-3pt}\raisebox{-15pt}[\headheight][0pt]{\vbox{\hbox to \textwidth
{\thepage \hfil \evenhead}\vskip4pt \hrule}}}
\renewcommand{\@oddhead}{
\hspace*{-3pt}\raisebox{-15pt}[\headheight][0pt]{\vbox{\hbox to \textwidth
{\oddhead \hfil \thepage}\vskip4pt\hrule}}}
\renewcommand{\@evenfoot}{}
\renewcommand{\@oddfoot}{}

\long\def\@makecaption#1#2{%
  \vskip\abovecaptionskip
  \sbox\@tempboxa{\small \textbf{#1.}\ \ #2}%
  \ifdim \wd\@tempboxa >\hsize
    {\small \textbf{#1.}\ \ #2}\par
  \else
    \global \@minipagefalse
    \hb@xt@\hsize{\hfil\box\@tempboxa\hfil}%
  \fi
  \vskip\belowcaptionskip}

\newcommand{\JNMPnumberwithin}[3][\arabic]{%
  \@ifundefined{c@#2}{\@nocounterr{#2}}{%
    \@ifundefined{c@#3}{\@nocnterr{#3}}{%
      \@addtoreset{#2}{#3}%
      \@xp\xdef\csname the#2\endcsname{%
        \@xp\@nx\csname the#3\endcsname .\@nx#1{#2}}}}%
}

\renewenvironment{proof}[1][\proofname]{\par
  \normalfont
  \topsep6\p@\@plus6\p@ \trivlist
  \item[\hskip\labelsep\textbf{%
    #1\@addpunct{.}}]\ignorespaces
}{%
  \qed\endtrivlist
}

\newcommand{\resetfootnoterule} {
  \renewcommand\footnoterule{%
  \kern-3\p@
  \hrule\@width.4\columnwidth
  \kern2.6\p@}
}

\renewcommand{\footnoterule}{}

\makeatother

\theoremstyle{definition}
\newtheorem*{definition}{Definition}

\begin{document}

\renewcommand{\evenhead}{ {\LARGE\textcolor{blue!10!black!40!green}{{\sf \ \ \ ]ocnmp[}}}\strut\hfill Jarmo Hietarinta}
\renewcommand{\oddhead}{ {\LARGE\textcolor{blue!10!black!40!green}{{\sf ]ocnmp[}}}\ \ \ \ \   Two-component versions of lattice equations}

\thispagestyle{empty}
\newcommand{\FistPageHead}[3]{
\begin{flushleft}
\raisebox{8mm}[0pt][0pt]
{\footnotesize \sf
\parbox{150mm}{{Open Communications in Nonlinear Mathematical Physics}\ \  \ {\LARGE\textcolor{blue!10!black!40!green}{]ocnmp[}}
\ \ Vol.2 (2022) pp
#2\hfill {\sc #3}}}\vspace{-13mm}
\end{flushleft}}

\FistPageHead{1}{\pageref{firstpage}--\pageref{lastpage}}{ \ \ Article}

\strut\hfill

\strut\hfill

\copyrightnote{The author. Distributed under a Creative Commons Attribution 4.0 International License}

\Name{Search for integrable two-component versions of the lattice
  equations in the ABS-list}

\Author{Jarmo Hietarinta}

\Address{Department of Physics and Astronomy\\
  University of Turku, FIN-20014 Turku, Finland\\{\tt hietarin@utu.fi}}

\Date{Received Date: July 16, 2022; Accepted Date: July 25, 2022}

\setcounter{equation}{0}

\begin{abstract}
\noindent 
We search and classify two-component versions of the quad
  equations in the ABS list, under certain assumptions. The
  independent variables will be called $y,z$ and in addition to
  multilinearity and irreducibility the equation pair is required to
  have the following specific properties: (1) The two equations
  forming the pair are related by $y\leftrightarrow z$ exchange. (2)
  When $z=y$ both equations reduce to one of the equations in the ABS
  list. (3) Evolution in any corner direction is by a multilinear
  equation pair. One straightforward way to construct such
  two-component pairs is by taking some particular equation in the ABS
  list (in terms of $y$), using replacement $y \leftrightarrow z$ for
  some particular shifts, after which the other equation of the pair
  is obtained by property (1). This way we can get 8 pairs for each starting
  equation. One of our main results is that due to condition (3) this is
  in fact complete for H1, H3, Q1, Q3. (For H2 we have a further case,
  Q2, Q4 we did not check.) As for the CAC integrability test, for
  each choice of the bottom equations we could in principle have $8^2$
  possible side-equations.  However, we find that only equations
  constructed with an {\em even} number of $y \leftrightarrow z$
  replacements are possible, and for each such equation there are two
  sets of ``side'' equation pairs that produce (the same) genuine
  B\"acklund transformation and Lax pair.
\end{abstract}

\label{firstpage}


\section{Introduction}
\newcommand{\bu}[1]{\accentset{\bullet}{#1}}

Within the topic of integrable discrete systems \cite{HJN}, equations
that can be defined on a single quadrilateral of the Cartesian
$\mathbb Z\times\mathbb Z$ lattice have been studied in great
detail. One common equation type is defined by the following:
\begin{definition} {\bf 1.} [Acceptable one-component quad equations]\ 
  \begin{enumerate}
  \item[{\bf1.1}] The equation depends on all corner variables of the elementary
  quadrilateral.
  
\item[{\bf1.2}] The equation is affine linear in each corner variable.
  
\item[{\bf1.3}] The equation is irreducible.
  
\item[{\bf1.4}] Uniformity: Every quadrilateral in the plane carries the same
  equation (depending on corresponding corner variables)
\end{enumerate}
\end{definition}

\begin{figure}
\centering
\begin{tikzpicture}[scale=2.0]
 \draw[thin] (-0.3,0) -- (1.3,0);
 \draw[thin] (-0.3,1) -- (1.3,1);
 \draw[thin] (0,-0.3) -- (0,1.3);
 \draw[thin] (1,-0.3) -- (1,1.3);
\draw [black] (1,1) circle (1.2pt);
\filldraw [black] (0,0) circle (1.2pt);
\filldraw [black] (1,0) circle (1.2pt);
\filldraw [black] (0,1) circle (1.2pt);
\node at (-0.3,0.16) {$u_{n,m}$};
\node at (1.36,0.16) {$u_{n+1,m}$};
\node at (0.5,0.16) {$p$};
\node at (-0.35,1.16) {$u_{n,m+1}$};
\node at (-0.15,0.56) {$q$};
\node at (1.46,1.16) {$u_{n+1,m+1}$};
\end{tikzpicture}
\caption{Corner variables on an elementary quadrilateral.\label{F:2}}
\end{figure}
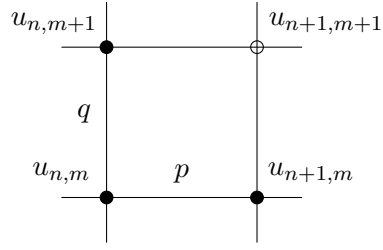

The geometric description is in Figure \ref{F:2}: subscript $m$ labels
the points in the vertical direction and $n$ in the horizontal
direction.  The lattice parameters $p,\,q$ are associated with
horizontal and vertical directions, respectively.
In practice we use shorthand notation in which a shift in the
$n$-direction is indicated by a tilde, and in the $m$-direction by a hat
\[
u_{n,m}=u,\quad u_{n+1,m}=\t u,\quad u_{n,m+1}=\h u,\quad
u_{n+1,m+1}=\xht u.
\]

 If the conditions in
Definition {\bf 1} are satisfied one can define evolution starting
from staircase- or corner-like initial conditions.

\begin{figure}
\centering
\begin{tikzpicture}[scale=1.5,tdplot_main_coords, cube/.style={thick,black,inner sep
=0pt}]
\draw[cube] (0,2,0) -- (2,2,0) -- (2,0,0);
\draw[cube] (0,0,2) -- (0,2,2) -- (2,2,2) -- (2,0,2) -- cycle;
        \draw[dashed] (0,0,0) -- (0,0,2);
        \draw[dashed] (0,0,0) -- (0,2,0); 
        \draw[dashed] (0,0,0) -- (2,0,0);
       \draw[cube] (0,2,0) -- (0,2,2);
        \draw[cube] (2,0,0) -- (2,0,2);
        \draw[cube] (2,2,0) -- (2,2,2);
\node at (0,0,-0.2) {$u$};
\node at (2,0,-0.2) {$\t u$};
\node at (2,2.1,-0.2) {$\xht u$};
\node at (0,2,-0.2) {$\h u$};
\node at (0,0.2,2.2) {$\xb u$};
\node at (2,0,2.3) {$\xbt u$};
\node at (2,2,2.3) {$\xbht u$};
\node at (0,2,2.3) {$\xbh u$};
\draw[thick,->] (2,0,0) -- (3,0,0);
\draw[thick,->] (0,2,0) -- (0,2.7,0);
\draw[thick,->] (0,0,2) -- (0,0,2.7);
\node at (0,0.4,2.7) {$r,\, k$};
\node at (3,-0.4,0) {$p,\, n$};
\node at (0,2.7,0.2) {$q,\, m$};
\draw[black,fill] (0,0,0) circle [radius=0.05cm];
\draw[black,fill] (0,0,2) circle [radius=0.05cm];
\draw[black,fill] (0,2,0) circle [radius=0.05cm];
\draw[black,fill] (2,0,0) circle [radius=0.05cm];

\draw[gray!50,fill] (2,2,0) circle [radius=0.05cm];
\draw[black] (2,2,0) circle [radius=0.051cm];
\draw[gray!50,fill] (2,0,2) circle [radius=0.05cm];
\draw[black] (2,0,2) circle [radius=0.051cm];
\draw[gray!50,fill] (0,2,2) circle [radius=0.05cm];
\draw[black] (0,2,2) circle [radius=0.051cm];

\draw[gray!10,fill] (2,2,2) circle [radius=0.07cm];
\draw[black] (2,2,2) circle [radius=0.071cm];

\end{tikzpicture}
\caption{The consistency cube.\label{F:3}}
  \end{figure}
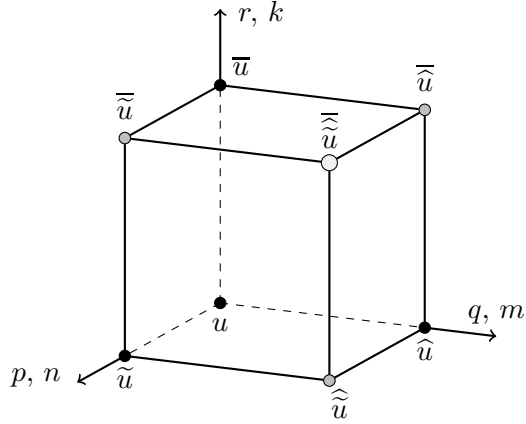

For lattice equations a necessary property for integrability is
``Multidimensional Consistency''. It means that the equations can be
{\em consistently} extended into higher dimensions, which is related
to the existence of a {\em hierarchy} of integrable continuous
equations (\cite{HJN}, Sec 3.2).  For 2D quad equations it means in
practice Consistency-Around-a-Cube (CAC), that is, the original quad
equation can be put on a 3D cube in a consistent way. Consider Figure
\ref{F:3} and assume that the original 2D lattice equation is on the
bottom of the cube. Certain modifications of that equation are then
placed on the back and left sides. Typically these equations are
obtained by cyclic permutation:
\begin{equation}\label{cycrule}
  \t {\ } \,\to\,\h {\ }\, \to\,\overline{\phantom{a}}\to \t {\ }\qquad
  p\,\to\,q\,\to\, r\,\to\, p
  \qquad  n\,\to\,m\,\to\, k\,\to\, n
\end{equation}
where we have also introduced a bar to denote shift in the vertical
direction, where steps are counted by $k$ : $u_{n,m,k+1}=\xb u$.  The
equations on the opposing sides are obtained by the perpendicular
shift. We then have 6 equations
\bse\label{eq:conseqs}\begin{align}
\text{bottom:}\quad & Q_{12}(u,\t u,\h u,\xht u;p,q)=0.&
\text{top:}\quad &Q_{12}(\xb u,\xbt u,\xbh u,\xbht u;p,q)=0,&\\ 
\text{back:}\quad & Q_{23}(u,\h u,\xb u,\xbh u;q,r)=0,&
\text{front:} \quad &Q_{23}(\t u,\xht u,\xbt u,\xbht u;q,r)=0,&\\
\text{left:}\quad & Q_{31}(u,\xb u,\t u,\xbt u;r,p)=0,&
\text{right:} \quad &Q_{31}(\h u,\xbh u,\xht u,\xbht u;r,p)=0.&
\end{align}
\ese

\tdplotsetmaincoords{75}{115}

The consistency problem arrives as follows: Take $u,\,\t u,\,\h
u,\,\xb u$ as initial values, then from bottom, back and left
equations we can compute the values of $\xht u$, $\xbh u$ and $\xbt
u$, respectively. After these are substituted into the top, front and
right equations we get independently 3 values for $\xbht u$ and these
values must be the same. This introduces severe conditions.

Several isolated examples of integrable quad-equations were found
already in the 1980s by considering continuous equations and the
permutability property of their B\"acklund transformations
(\cite{HJN}, Sec. 2.4-5). A major development in this field was the
classification of integrable quad-equations by Adler, Bobenko and
Suris \cite{ABS03}, under the assumptions of D4 symmetry and the
``tetrahedron property''. (The tetrahedron property was essential in
the classification work. It states that the triply shifted quantity
computed in three ways from \eqref{eq:conseqs} does not depend on
$u$.)  The result of this classification is the so-called
``ABS-list'', its main components being the H and Q lists:

\vskip 0.4cm
{\bf H-list}
\bse\label{eq:Heqs}\begin{eqnarray}
&&{\rm H_1}: \quad (u-\wh{\wt{u}})(\wh{u}-\wt{u})=p^2-q^2  \\ 
  &&{\rm H_2}: \quad (u-\wh{\wt{u}})(\wt{u}-\wh{u})=
  (p-q)(u+\wt{u}+\wh{u}+\wh{\wt{u}})+p^2-q^2  \\ 
  &&{\rm H_3}: \quad p(u\wt{u}+\wh{u}\wh{\wt{u}})-
  q(u\wh{u}+\wt{u}\wh{\wt{u}})=\dd^2(p^2-q^2) 
\end{eqnarray}\ese

 {\bf Q-list}:
\noindent
\nopagebreak
\bse\label{eq:Qeqs}\begin{eqnarray}
  &&{\rm Q_1}: \quad  p(u-\wh{u})(\wt{u}-\wh{\wt{u}})-
  q(u-\wt{u})(\wh{u}-\wh{\wt{u}})=\dd^2pq(q-p) \\ 
  &&{\rm Q_2}: \quad  p(u-\wh{u})(\wt{u}-\wh{\wt{u}})-
  q(u-\wt{u})(\wh{u}-\wh{\wt{u}})+pq(p-q)(u+\wt{u}+\wh{u}+\wh{\wt{u}})\nn \\ 
  && \hspace{3cm}  =pq(p-q)(p^2-pq+q^2) \\ 
  &&{\rm Q_3}: \quad p(1-q^2)(u\wh{u}+\wt{u}\wh{\wt{u}})-
  q(1-p^2)(u\wt{u}+\wh{u}\wh{\wt{u}}) \nn \\ 
  && \hspace{3cm} =(p^2-q^2)\left((\wh{u}\wt{u}+u\wh{\wt{u}})+
  \dd^2\frac{(1-p^2)(1-q^2)}{4pq}\right)\label{eq:Q3inABS}  \\
  &&{\rm Q_4}\text{ from \cite{Hie05}}:
\quad
\sn(\ar)(u\wt{u}+\wh{u}\wh{\wt{u}})
-\sn(\bb)(u\wh{u}+\wt{u}\wh{\wt{u}}) 
-\sn(\ar-\bb)(\wt{u}\wh{u}+u\wh{\wt{u}}) \nn \\ 
  && \hspace{3cm} +k\,\sn(\ar)\sn(\bb)\sn(\ar-\bb)
(1+u\wt{u}\wh{u}\wh{\wt{u}})=0.
\end{eqnarray}\ese 

However, it is well  known that there are other CAC-compatible
equations if some conditions used by ABS are relaxed, for example
$\xht u\, u -\t u\,\h u=0$, which breaks the tetrahedron condition.

One of the assumptions used to generate the ABS list was that
equations on opposing sides are related by the corresponding shift, as
can be seen in \eqref{eq:conseqs}. This assumption was relaxed in the
work of Boll \cite{Boll2011,Boll2012JNMP}, while still
keeping the tetrahedron property. On the other hand, in the
classification of Hietarinta \cite{JHetsiquad} the tetrahedron
assumption was not made but the search was restricted to equations
that were quadratic homogeneous.

When we have a set of consistent equations on the sides of the cube
one can use the ``side'' equations to construct a Lax pair or a
B\"acklund transformation, which should generate the ``bottom''
equation (see e.g., \cite{HJN}, Sec. 3.3). But in \cite{JHetsiquad} it
was found that many equations can pass the CAC test without being
integrable, in other words, sometimes the Lax pair generated from the
side equations is trivial. This means that CAC is only a necessary
test and must be verified by the existence of a genuine Lax or
B\"acklund pair.

\section{Previous work on two-component equations}
Multi-component quad equations have also been studied and various
types of equations have been proposed. For example discrete versions
of the Boussinesq equations have been proposed, often in
three-component form \cite{H-JPA-2011}, but after eliminating one
variable one obtains in some cases a two component form still on the
elementary quadrilateral (e.g., \cite{N99}, (4.8)). Several
two-component equations were also proposed in \cite{FX17}. However,
none of these equations satisfy the exchange conditions {\bf
  2.2} in Definition {\bf 2} below.

Furthermore in this paper we restrict our attention to equations that
can be considered as multi-component generalizations of the equations
in the ABS-list. One such equation was given in \cite{BHQK} (table 5,
with name change $x\to y,\,y\to z$)
\begin{equation}\label{eq:here} \left\{ \begin{array}{r}
    (y- \xht y)(\t z-\h z)-p^2+q^2=0,\\
  (z- \xht z)(\t y-\h y)-p^2+q^2=0.\end{array}\right.
\end{equation}
Clearly the limit $z \to y$ (for any shift: none, tilde, hat,
tilde-hat) takes both equations to H1.  Furthermore the equations are
related by $y \leftrightarrow z$ exchange for all shifts.  Also
note that in the first equation the {\em once shifted} variables have
been changed by $y\to z$.

Is this equation integrable? At least it should have the CAC property.
With \eqref{eq:here} as the bottom equation we have to choose the side
equations. It would be natural to try the cyclic rule \eqref{cycrule},
which produces \bse\label{cyccomb}\begin{align}\text{bottom:}\hskip
  1cm &\left\{ \begin{array}{r} (y- \xht y)(\t z-\h z)+q-p=0,\\ (z-
    \xht z)(\t y-\h y)+q-p=0,
  \end{array}\right.\\
  \text{back:}\hskip 1cm &\left\{ \begin{array}{r}
    (y- \xbh y)(\h z-\xb z)+r-q=0,\\
  (z- \xbh z)(\h y-\xb y)+r-q=0,
    \end{array}\right.\\
  \text{left:}\hskip 1cm &\left\{ \begin{array}{r}
    (y- \xbt y)(\xb z-\t z)+p-r=0,\\
    (z- \xbt z)(\xb y-\t y)+p-r=0.\end{array}\right.
\end{align}\ese
This indeed passes the CAC test with the triply shifted variables being
\bse\begin{eqnarray}
\xbht y=\frac{ p \t y (\xb y - \h y) + q \h y ( - \xb y + \t y) + r
  \xb y (\h y - \t y)}{p (\xb y - \h y) + q ( - \xb y + \t y) + r (\h
  y - \t y)},\\ \xbht z=\frac{ p \t z (\xb z - \h z) + q \h z ( - \xb z
  + \t z) + r \xb z (\h z - \t z)}{p (\xb z - \h z) + q ( - \xb z + \t
  z) + r (\h z - \t z)}.
\end{eqnarray}\ese
Note that this has the tetrahedron property (no unshifted $y,z$) and
that the $y$ and $z$ variables are separated in the final formulae.

The above construct in which variables with an odd number of shifts
are exchanged, was described already in \cite{AtkThesis} as Toeplitz
extension. It was generalized to all the equations in the ABS list in
\cite{FZZ-CPL} by the same rule: exchanging the singly shifted
variables, see also \cite{KNPT}. This approach was developed further
by including other replacements by symmetry arguments \cite{ZvdKZ}.
Another approach in deriving multi-component versions of the ABS list
was given in \cite{Kels} where such equations were derived by from the
star-triangle relations.

One result in \cite{ZvdKZ} was the following two-component version of
H1 (Eqs.\ (2.25), (2.28),(2.29)) consisting of
\bse\label{eq:ZvdKZ} \begin{align}\text{bottom:}\hskip 1cm
  &\left\{ \begin{array}{r} (z- \xht y)(\t z-\h y)+q-p=0,\\ (y- \xht
    z)(\t y-\h z)+q-p=0,
  \end{array}\right.\label{eq:ZvdKZa}\\
  \text{back:}\hskip 1cm &\left\{ \begin{array}{r}
    (z- \xbh y)(\h y-\xb z)+r-q=0,\\
  (y- \xbh z)(\h z-\xb y)+r-q=0,
    \end{array}\right.\label{eq:ZvdKZb}\\
  \text{left:}\hskip 1cm &\left\{ \begin{array}{r}
    (y- \xbt y)(\xb y-\t y)+p-r=0,\\
    (z- \xbt z)(\xb z-\t z)+p-r=0,\end{array}\right.\label{eq:ZvdKZc}
\end{align}\ese
Here the bottom and back equations are related by cyclic permutation,
but the left equation is of entirely different type, in fact
separating the $y$ and $z$ variables. This peculiar combination passes
the CAC test, and the triply shifted variables are
\bse\begin{align}
  \xbht y&=\frac{p \t z (\h y - \xb z) + q \h y (\xb z - \t z) + r \xb z ( - \h
    y + \t z)} {p (\h y - \xb z) + q (\xb z - \t z) + r ( - \h y +
    \t z)},\\ \xbht z&=\frac{p \t y (\xb y - \h z) + q \h z ( - \xb y + \t y) +
    r \xb y ( - \t y + \h z) }{ p (\xb y - \h z) + q ( - \xb y + \t y) + r ( -
    \t y + \h z)},
\end{align}\ese
and they have the tetrahedron property.

\section{Classification of two-component generalizations of the ABS list}
The puzzling triplet \eqref{eq:ZvdKZ} suggests that there may be
interesting phenomena specific for two-component equations. The
purpose of this paper is to search and classify such equations.

\subsection{The domain of the search}
Since the fully generic case of the problem is too hard to tackle we
restrict our attention to equation pairs with the following
properties:
\begin{definition}{\bf 2}\ [Acceptable two-component quad equations]\ 
 \begin{enumerate}
\item[{\bf 2.1}] Both equations of the pair are affine multilinear and
  irreducible.
\item[{\bf 2.2}] Exchange rule: The two equations that form the pair
  are related by the exchange rule $y \leftrightarrow z,\, \t y
  \leftrightarrow \t z,\, \h y \leftrightarrow \h z,\,\xht y
  \leftrightarrow \xht z$.
\item[{\bf 2.3}] Evolution: From the pair of equations one can solve
  for any of the corner variable pairs $\{y,z\}, \{\t y,\t z\},\, \{\h
  y,\h z\},\, \{\xht y,\xht z\}$
\item[{\bf 2.4}] Strong multilinearity: When any resolved variable pair is
  written as a pair of polynomial equations, the polynomials are again
  multilinear and irreducible.
\end{enumerate}
\end{definition}

\paragraph{Remarks:}
\begin{itemize}
  \item We use $\bullet$ to indicate when the exchange rule {\bf 2.2}
    has been applied, That is, if $B$ is obtained from $A$ by the
    exchange rule we write $B=\bu A$. Obviously $({\bu
      A}\,)\!\bu{\phantom A}=A$.
  \item Multilinearity does not imply unique evolution. Consider the pair
    \begin{eqnarray*}
z \xht y + \h y \h z + 2 \h y \t z + \t z \t y&=&0,\\ y \xht z + \h y
\h z + 2 \h z \t y + \t z \t y&=&0.
\end{eqnarray*}
    As given it is resolved for $\{\xht y,\xht z\}$ and for
    $\{y,z\}$. However, if one tries solve for $\{\t y,\t z\}$ or
    $\{\h y,\h z\}$ there will be square roots and therefore evolution
    in the NW or SE direction is not uniquely determined. Note also
    that the one-component reduction of this pair is not multilinear.
  \item Multilinearity does not imply strong multilinearity.  Consider
    the pair of equations
    \begin{eqnarray*}
    (2y-z-\xht y)(2\h z -\h y-2\t z+\t y)&=&p^2-q^2,\\
      (2z-y-\xht z)(2\h y -\h z-2\t y+\t z)&=&p^2-q^2,
      \end{eqnarray*}
    which is resolved for $\xht y,\xht z$. It is obviously multilinear
    and reduces to H1. When this pair is solved for $\{\t y,\t z\}$
    one obtains the pair
     \begin{eqnarray*}
       3 (\h y-\t y) (2 y - \xht y - z) (y - 2 z + \xht z)&=&(p^2-q^2)
       (2 \xht y + \xht z  - 3 y),\\ 3 (\h z-\t z) (2 z - \xht z - y) (z - 2 y +
     \xht y)&=&(p^2-q^2) (2 \xht z + \xht y - 3 z),
     \end{eqnarray*}
     which is resolved for both $\{\t y,\t z\}$ and $\{\h y,\h z\}$.
     However, this is not multilinear because $y$ and $z$ appear
     quadratically. (As a consequence, if we attempt to resolve for
     $y,z$ from this pair there is a superfluous solution.)
\end{itemize}

\subsection{Results for acceptable pairs}
\begin{prop}\label{P1}
  1. For any given equation in the ABS list one can get a two
  component version satisfying the conditions in Definition {\bf 2}
  by applying to the original equation any one of the following eight
  replacements \bse \label{subs} \begin{eqnarray} 0:&&
    \text{none},\\ 1:&& y \rightarrow z,\\ 2:&& \t y \rightarrow \t
    z,\\ 3:&& \h y \rightarrow \h z,\\ 4:&& y \rightarrow z,\,\t y
    \rightarrow \t z,\\ 5:&& y \rightarrow z,\,\h y \rightarrow \h
    z,\\ 6:&& \t y \rightarrow \t z,\,\h y \rightarrow \h z,\\ 7:&& y
    \rightarrow z,\,\t y \rightarrow \t z,\, \h y \rightarrow \h z,\,
  \end{eqnarray}    \ese
after which the other member of the pair is
obtained by the exchange rule {\bf 2.2}

2. For H1, H3, Q1 and Q3 this result is complete.
  \end{prop}

For H2 we have a counterexample on completeness, given below, while
for Q2 and Q4 uniqueness is open.

\begin{proof}
  1. It is easy to verify that from an equation in the ABS list, any
  of the substitutions \eqref{subs} results in a pair satisfying all
  properties of Definition {\bf 2}.

  2. It is a bit more laborious to show that there are no others.  For
  this purpose we generate multilinear equations (with arbitrary
  coefficients) for all four resolutions, i.e. equation pairs of the
  type \bse\label{eq:pairset} \begin{align}
    &\left\{ \begin{array}{r}\xht y\, L_1(y,z,\t y,\t z,\h y,\h
      z)+P_1(y,z,\t y,\t z,\h y,\h z)+C=0,\label{eq:setyht}\\ \xht z\,
      L_1(z,y,\t z,\t y,\h z,\h y)+P_1(z,y,\t z,\t y,\h z,\h
      y)+C=0,\end{array}\right.  \\ &\left\{ \begin{array}{r}\t y\,
      L_2(y,z,\h y,\h z,\xht y,\xht z)+P_2(y,z,\h y,\h z,\xht y,\xht
      z)+C=0,\\ \t z\, L_2(z,y,\h z,\h y,\xht z,\xht y)+P_2(z,y,\h
      z,\h y,\xht z,\xht y)+C=0,\end{array}\right.
    \\ &\left\{ \begin{array}{r}\h y\, L_3(y,z,\t y,\t z,\xht y,\xht
      z)+P_3(y,z,\t y,\t z,\xht y,\xht z)+C=0,\\ \h z\, L_3(z,y,\t
      z,\t y,\xht z,\xht y)+P_3(z,y,\t z,\t y,\xht z,\xht
      y)+C=0,\end{array}\right.  \\ &\left\{ \begin{array}{r}y\,
      L_4(\t y,\t z,\h y,\h z,\xht y,\xht z)+P_4(\t y,\t z,\h y,\h
      z,\xht y,\xht z)+C=0,\\ z\, L_4(\t z,\t y,\h z,\h y,\xht z,\xht
      y)+P_4(\t z,\t y,\h z,\h y,\xht z,\xht y)+C=0,\end{array}\right.
    \end{align}\ese
where $L_j$ are linear and $P_j$ quadratic multilinear polynomials in
the indicated variables.  Next some coefficients in $L_j,\,P_j$ are
fixed by the condition that the $z\mapsto y$ reduction leads to one of
the equations H1, H3, Q1, Q3. This still leaves 3 free coefficients
in each $L_j$ and 9 in $P_j$.  The pairs in \eqref{eq:pairset}
describe the same evolution and therefore if we solve $\{\xht y,\xht
z\}$ from \eqref{eq:setyht}, say, and substitute to the other
equations they should all vanish. This leads to 384 smallish
equations, which can be solved by starting with the simplest ones and
proceeding step by step. This is not difficult, only tedious. For each
of the equations H1, H3, Q1, Q3 the solution process eventually splits
into eight branches as listed in \eqref{subs}.  \end{proof}

\subsubsection{H2}
For H2 we found an equation that does not fit into the result of
Proposition \ref{P1}:
\begin{align}
&(\xht y-(y+z)/2) (\h y - \t y + \h z - \t z)\nn\\
&+\nu_1\, (y + \t y + z + \t z+2\,\epsilon\, p) (\h y - \h z)\nn\\
&+\nu_2\, (y + \h y + z + \h z+2\,\epsilon\, q) (\t y - \t z)\nn\\
&+\nu_3\, (\h y - \t y + \h z - \t z-2\,\epsilon\, (p-q)) (y-z)\nn\\
&-\epsilon\, (p-q)(2\,\xht y+y+z + \h y + \t y + \h z + \t z)\nn\\
&-2 \epsilon^2 (p^2-q^2)=0,\label{eq:H2x}
\end{align}
together with its $z\leftrightarrow y$ reflection.  This pair
satisfies the strong multilinearity condition if all parameters $\nu_i$
are nonzero.  It reduces to H2 when all $z=y$, and clearly the
parameters $\nu_j$ disappear in this reduction. We do not know whether
\eqref{eq:H2x} is integrable or linearizable.

\section{Integrability}
\subsection{Integrability by CAC}
The example \eqref{cyccomb} shows that if one uses the replacement
rule 6 and its cyclic variants for the bottom back and left equations
(which we denote as $(6,6,6)$) the system has CAC property. On the
other hand in  example \eqref{eq:ZvdKZ} the replacements are given by
$(4,5,0)$. The question then arises as to which combinations among the
$8^3$ possibilities have the CAC property. The result is as follows:
\begin{prop}\label{P2}
  For each equation in the ABS list the following eight replacement
  rules have the CAC property: (0,0,0), (0,4,5), (4,5,0), (4,6,5),
  (5,0,4), (5,4,6), (6,5,4), (6,6,6), where the numbers in the triplet
  are the replacement rules used for bottom, back, and left equations
  on the cube. The rules given in \eqref{subs} must be modified
  cyclically to fit the corresponding side. The top, front and right
  equations are obtained by a perpendicular shift.
\end{prop}

\begin{proof} By direct computation. Since there are no free
  parameters the computations for the 512 cases are easy to automatize.
  \end{proof}

Remarks:
\begin{itemize}
  \item The only replacements appearing in the list are 0,4,5,6, which
    correspond to replacements of even number of variables.
  \item The cases (0,4,5), (4,5,0), (5,0,4), are related by rotation
    around the $(y,z) - (\xbht y,\xbht z)$ axis, the same holds for
    (4,6,5), (6,5,4), (5,4,6). There are therefore only four
    essentially different triplets.
  \item The fact that there are two kinds of side equation pairs for each
    bottom equation pair follows from the $y\leftrightarrow z$
    symmetry. For if we do this exchange only on the variables
    with a bar-shift, the top pair does not change (and neither does
    the bottom pair), but the side equations will change.
   \item Our end result agrees with the result of \cite{ZvdKZ}, which
     was derived by an entirely different approach.
\end{itemize}    


\subsection{Decoupling}
It is easy to see that in the (0,0,0) case the equations are
decoupled, since in each pair one equation depends only on the $y$
variables and the other only on the $z$ variables. We will now look
whether the other sets can also be decoupled somehow.

Since the CAC analysis is completely algebraic the variable names do
not matter: instead of $y, z, \t y, \t z, \h y, \dots$ we could have
used $a, b, c, d, \dots$ and the algebra would have been the same.

In the case of (4,5,0) given in \eqref{eq:ZvdKZ} we see that
\eqref{eq:ZvdKZc} is already decoupled and we can separate variables
into two set, $S_a= \{y, \t y, \xb y,\xbt y\}$ and $S_b= \{z, \t z,
\xb z,\xbt z\}$.  Insisting that any particular equation depends only
on variables from one set, we can augment these sets using
\eqref{eq:ZvdKZ} and its shifted version to
\bse\begin{align}\label{eq:sets450} (4,5,0):& & S_a= \{y, \t y, \xb
y,\xbt y,\h z,\xbh z, \xht z, \xbht z\},\quad S_b= \{z, \t z, \xb
z,\xbt z,\h y,\xbh y, \xht y,\xbht y\}.
\end{align}
Thus there are 6 equations depending on variables from $S_a$ and they
satisfy CAC all be themselves, similarly for $S_b$.  It seems that all
equations that pass the CAC test do decouple in the described
manner. For example while (6,5,4) passes CAC and decouples, (6,4,5)
does not decouple nor pass CAC. The sets for the other integrable
combinations are as follows:
\begin{align}\label{eq:sets000}
(0,0,0):& &  S_a= \{y, \t y, \xb y,\xbt y,\h y,\xbh y, \xht y, \xbht y\},\quad
  S_b= \{z, \t z, \xb z,\xbt z,\h z,\xbh z, \xht z,\xbht z\},\\
(6,5,4):& &  S_a= \{y, \t z, \xb y,\xbt z,\h z,\xbh z, \xht y, \xbht y\},\quad
  S_b= \{z, \t y, \xb z,\xbt y,\h y,\xbh y, \xht z,\xbht z\},
  \label{eq:sets654}\\
(6,6,6):& &  S_a= \{y, \t z, \xb z,\xbt y,\h z,\xbh y, \xht y, \xbht z\},\quad
  S_b= \{z, \t y, \xb y,\xbt z,\h y,\xbh z, \xht z,\xbht y\}.
  \label{eq:sets666}
\end{align}
\ese These can be described as follows: (0,0,0) is decoupled as it
stands; (4,5,0) is decoupled if for odd number of hat-shits we
exchange $z\leftrightarrow y$; (6,5,4) can be decoupled if for odd
total number of tilde and hat-shifts we exchange, while bar shift has
no effect; for (6,6,6) exchange is needed when the total number of all
kinds of shifts is odd.

If the exchanges needed for decoupling are transferred to a property
of the lattice itself, then for (0,0,0) the lattice is uniform; for
(4,5,0) we should change the lattice on alternate planes in the hat
direction; for (6,5,4) we have checkerboard lattice in tilde and hat
direction without change in bar direction; for (6,6,6) we need a
lattice that is alternating in every direction.

The possibility of decoupling follows in part from our assumption that
the two equations are related by $y\leftrightarrow z$ replacement. But
there are other two-component equation pairs for which decoupling is
not possible, for example the discrete Boussinesq equation given
e.g.~as the pair (3.3) of \cite{rev} has $\t w,\t z,\h w,\h z$ in both
equations of the pair.

The above decoupling approach explains the algebra quite well, but in
practice the variables are not featureless symbols but the
tilde-hat-bar decorations have a dynamic meaning. That is, if we have
a solution $y=y(n,m,k),\,z=z(n,m,k)$ for the set of equations then $\h
y,\,\h z$ are obtained by changing $m\to m+1$ in the concrete formula
for $y$ and $z$. If the $y$ and $z$ solutions are different (for example
solitons traveling in different directions) it may be better to keep
the dependent variables uncoupled and the equations coupled than vice
versa.


\subsection{Integrability by BT}
It has been observed that CAC is necessary but not always sufficient
for integrability \cite{JHetsiquad}. However, the existence of a
genuine B\"acklund Transformation (BT) (or equivalently, a nontrivial
Lax pair) is a proof of integrability.

\begin{prop}
  For each equation in the ABS list the following eight replacement
  rules (0,0,0), (0,4,5), (4,5,0), (5,0,4), (4,6,5), (6,5,4), (5,4,6),
  (6,6,6), generate a genuine BT. That is, if one assigns any of the
  listed triplet replacements on the bottom, back, and left pairs of
  equations (or their cyclic permutations), then one can freely choose
  two pairs of ``side'' equations for BT and together with their
  perpendicular shift equations they generate by variable elimination
  the third pair.
\end{prop}

For example if the bottom pair is generated by replacement 5, back
pair by 4 then they together generate left pair of type 6, after three
of the four variables on the right pair are eliminated. The same left
pair is also generated if bottom and back equations are both generated
by replacement 6.

\subsection{The Lax pair}
One of the more important proofs of integrability is by construction
of a Lax pair. But here one must note that there are ``fake'' Lax
pairs \cite{BH13,JHetsiquad} and therefore one must verify that the
Lax pair is genuine and able to generate the equation(s) in
question. The general formula for constructing Lax pairs from a CAC
consistent system is given e.g., in \cite{HJN} Section 3.3.1, and
furthermore there are now even computer programs that can do that
\cite{BHQK,Bri}.

It is perhaps sufficient to consider an example, for which we choose
type 5 Q1. As noted before type 5 bottom equation goes together with
side equations 4,6 as well as 0,4.

\subsubsection{Example: Q1, sides 4,6 generate bottom 5}
We construct the Lax pair for Q1;5 from a back equation pair
of type 4 \bse\label{eq:q1ex}
\begin{eqnarray}
  q(z-\wb{y})(\wh{z}-\wb{\wh{y}})-
  r(z-\wh{z})(\wb{y}-\wb{\wh{y}})=\dd^2qr(r-q),\\
  q(y-\wb{z})(\wh{y}-\wb{\wh{z}})-
  r(y-\wh{y})(\wb{z}-\wb{\wh{z}})=\dd^2qr(r-q),
  \end{eqnarray}
and a left equation of type 6
\begin{eqnarray}
  r(y-\wt{z})(\wb{z}-\wb{\wt{y}})-
  p(y-\wb{z})(\wt{z}-\wb{\wt{y}})=\dd^2rp(p-r),\\
  r(z-\wt{y})(\wb{y}-\wb{\wt{z}})-
  p(z-\wb{y})(\wt{y}-\wb{\wt{z}})=\dd^2rp(p-r),
\end{eqnarray}
Together they should generate a bottom equation of
type 5, i.e.,
\begin{eqnarray}
  p(z-\wh{z})(\wt{y}-\wh{\wt{y}})-
  q(z-\wt{y})(\wh{z}-\wh{\wt{y}})=\dd^2pq(q-p),\\
  p(y-\wh{y})(\wt{z}-\wh{\wt{z}})-
  q(y-\wt{z})(\wh{y}-\wh{\wt{z}})=\dd^2pq(q-p).
\end{eqnarray}\ese

In order to construct the Lax pair we solve the first 4 equations of
\eqref{eq:q1ex} for the double shifted quantities and then replace the
barred quantities as follows:
\[
\wb y=\frac{f}{k},\quad
\wb z=\frac{g}{l},\quad
\wb{\wt y}=\frac{\wt f}{\wt k},\quad
\wb{\wt z}=\frac{\wt g}{\wt l},\quad
\wb{\wh y}=\frac{\wh f}{\wh k},\quad
\wb{\wh z}=\frac{\wh g}{\wh l}.
\]
For the left equations this leads to
\bse\begin{eqnarray}
  \frac{\wt f}{\wt k}&=&\frac{g[p\wt z+r(y-\wt z)]+pl[\delta^2r(r-p)-y\wt z]}
       {pg+l[-py+r(y-\wt z)]},\\
  \frac{\wt g}{\wt g}&=&\frac{f[p\wt y+r(z-\wt y)]+pk[\delta^2r(r-p)-z\wt y]}
       {pf+k[-pz+r(z-\wt y)]}.
\end{eqnarray}\ese
These can be written in matrix form 
\bse\begin{align}
  \wt{\psi}&={\cal L}_{Q1;6}\,\psi\quad \text{where}\quad \psi=(f,k,g,l)^T
  \quad \text{and}\quad {\cal L}_{Q1;6}=\begin{pmatrix}\mathbf 0 &\mathbf
  L\\ \bu{\mathbf L} & \mathbf 0\end{pmatrix}\quad\text{with}\\
{\mathbf L}&= \lambda\begin{pmatrix} p\wt z+r(y-\wt z) & p
   [\delta^2r(r-p)-y\wt z]\\ p & -py+r(y-\wt z)
 \end{pmatrix},\label{eq:Llaxdef}
  \end{align}\ese
where the parameter $\lambda$ is the splitting factor (and may depend
on $y,\wt z$ and therefore it is possible that $\lambda\neq \bu \lambda$).

Similarly, from the back equation we get
\bse\begin{align}
&\wh{\psi}= {\cal M}_{Q1;4}\,\psi\quad\text{where}\quad
   {\cal M}_{Q1;4}=\begin{pmatrix}\mathbf M & \mathbf 0\\ \mathbf 0 &
   \bu{\mathbf M }\end{pmatrix}\quad\text{with}
\\ &\mathbf M =\kappa\begin{pmatrix}
q\wh z+r(z-\wh z)] & q[\delta^2r(r-q)-z\wh z]\\
q & -qz+r(z-\wh z)]
  \end{pmatrix}.\label{eq:Mlaxdef}
\end{align}\ese

The commutativity condition $\wh{\cal L}{\cal M}=\wt{\cal M}{\cal L}$
now implies
\[
\wh{\mathbf L}\,\bu{\mathbf M}=\wt{\mathbf M}\,{\mathbf L},\quad\text{
or equivalently}\quad \bu{\wh{\mathbf L}}\,{\mathbf M}=\bu{\wt{\mathbf
      M}}\,\bu{\mathbf L}.
\]
In order to get the equations from this we have to fix the separations
constants $\lambda,\,\kappa$.  One elegant way to do that is to
require $\det {\cal L=\det M}=1$. Here it leads to \bse\begin{align}
\lambda^2=&1/[(\delta^2p^2-(y-\wt z)^2)(p-r)r],\quad
       {\bu\lambda}^2=1/[(\delta^2p^2-(z-\wt y)^2)(p-r)r],\\ \kappa^2
       =&1/[(\delta^2q^2-(y-\wh y)^2)(q-r)r],\quad
       \bu{\kappa}^2=1/[(\delta^2p^2-(z-\wh z)^2)(q-r)r].
\end{align}\ese
From these one can relatively easily derive
$(\wt{\kappa}\bu{\kappa}\,q)^2=(\lambda\wh \lambda p)^2$, but in practice
we need 
\begin{equation}\label{eq:factorq3}
\wt{\kappa}\bu{\kappa}\,q=\lambda\wh \lambda p,
\end{equation}
and its exchanged version, in order to derive the bottom
$Q1;5$-equation. (The apparent asymmetry in \eqref{eq:factorq3} is due to the
block structure of the Lax matrices).

\subsubsection{Example: Q1, sides 0,4 generate bottom 5}
A back equation pair of type 0 is given by  \bse\label{eq:q1exa}
\begin{eqnarray}
  q(y-\wb{y})(\wh{y}-\wb{\wh{y}})-
  r(y-\wh{y})(\wb{y}-\wb{\wh{y}})=\dd^2qr(r-q),\\
  q(z-\wb{z})(\wh{z}-\wb{\wh{z}})-
  r(z-\wh{z})(\wb{z}-\wb{\wh{z}})=\dd^2qr(r-q),
  \end{eqnarray}
and a left equation of type 4
\begin{eqnarray}
  r(z-\wt{y})(\wb{z}-\wb{\wt{y}})-
  p(z-\wb{z})(\wt{y}-\wb{\wt{y}})=\dd^2rp(p-r),\\
  r(y-\wt{z})(\wb{y}-\wb{\wt{z}})-
  p(y-\wb{y})(\wt{z}-\wb{\wt{z}})=\dd^2rp(p-r).
\end{eqnarray}\ese
Now comparing equations \eqref{eq:q1ex} and \eqref{eq:q1exa} we find
that the sets are the same if we exchange all barred quantities and
only them: $\wb y \leftrightarrow \wb z,\, \wb{\wt y} \leftrightarrow
\wb{\wt z},\, \wb{\wh y} \leftrightarrow \wb{\wh z}$. From the Lax
matrix point of view this means permuting the blocks, i.e.,
\[
 {\cal L}_{Q1;0}=\begin{pmatrix}\mathbf 0 &\bu{\mathbf
   L}\\ {\mathbf L} & \mathbf 0\end{pmatrix},\quad
  {\cal M}_{Q1;5}=\begin{pmatrix}\bu{\mathbf M} & \mathbf 0\\ \mathbf 0 &
  {\mathbf M }\end{pmatrix},
  \]
  while keeping the previous definitions \eqref{eq:Llaxdef}
  and  \eqref{eq:Mlaxdef}. Thus we end up with the same conditions.

\section{Conclusions}
In this paper we have searched for two-component versions of the
equations in the ABS list.  In addition to the standard assumptions
placed on quad equations we assumed the following (the dependent
variables are named $y,\,z$)
\begin{enumerate}
  \item The two equations forming the pair are related by
    $y\leftrightarrow z$ exchange (for all shifts).
  \item When $z=y$ both equations reduce to one of the equations in the
    ABS list.
  \item Evolution in any corner direction is by a pair of multilinear
    equations.
\end{enumerate}
Condition 3 in more detail: one must be able to solve for any corner
variable pair (e.g., $\wt y=A/B,\,\wt z=C/D$) and when this is written
as equations (e.g., $B\wt y-A=0,\,D\wt z-C=0$) they must be
multilinear in all of the dependent variables. We call this {\em
  strong multilinearity}.

The above conditions turn out to be quite strong and as a result we
found that the only possibility is that the pair of equations is
obtained from the original one component equation by a simple
replacement (Proposition \ref{P1}), the only caveats are H2, for which
we have a counterexample, and Q2 and Q4 which we did not check.

Since one can get several candidate pairs for each original equation
there rises a question related to multidimensional consistency, namely
how we should populate the sides of the consistency cube. We found
eight combinations that satisfy the CAC-condition, as described in
Proposition \ref{P2}. Our end result is essentially the same as the one
obtained in \cite{ZvdKZ} by symmetry arguments. Note also that if the
CAC condition is satisfied for some bottom equation, it is satisfied
with two different sets of side equations. Also, both pairs of
side equations work as a B\"acklund transformations generating the
same bottom equation, and from both pairs one gets the same Lax pair.
We gave the details for Q1 of type 5.

With the above equations one can ask some natural questions which
include: what are their semi-continuous and fully continuous limits,
and what are their soliton solutions, in particular how do the
different components of a soliton solution interact.

\subsection*{Acknowledgement} I would like to thank D-J Zhang for comments.
 All computations were done with REDUCE \cite{Hearn}.

\label{lastpage}
\end{document}